\documentstyle[12pt]{article}
\catcode`\@=11

\def\@normalsize{\@setsize\normalsize{15pt}\xiipt\@xiipt
\abovedisplayskip 14pt plus3pt minus3pt%
\belowdisplayskip \abovedisplayskip
\abovedisplayshortskip  \z@ plus3pt%
\belowdisplayshortskip  7pt plus3.5pt minus0pt}
\def\small{\@setsize\small{13.6pt}\xipt\@xipt
\abovedisplayskip 13pt plus3pt minus3pt%
\belowdisplayskip \abovedisplayskip
\abovedisplayshortskip  \z@ plus3pt%
\belowdisplayshortskip  7pt plus3.5pt minus0pt
\def\@listi{\parsep 4.5pt plus 2pt minus 1pt
            \itemsep \parsep
            \topsep 9pt plus 3pt minus 3pt}}

\def\underline#1{\relax\ifmmode\@@underline#1\else
        $\@@underline{\hbox{#1}}$\relax\fi}
\@twosidetrue
\relax

\catcode`@=12

\catcode`\@=11

\def\section{\@startsection{section}{1}{\z@}{3.5ex plus 1ex minus
   .2ex}{2.3ex plus .2ex}{\large\bf}}

\def\ps@headings{\def\@oddfoot{}\def\@evenfoot{}
\def\@oddhead{\hbox{}\hfill
        \makebox[.5\textwidth]{\raggedright\ignorespaces --\thepage{}--
        \hfill }}
\def\@evenhead{\@oddhead}
\def\subsectionmark##1{\markboth{##1}{}}
}

\catcode`\@=12

\relax

\def\figcap{\section*{Figure Captions\markboth
        {FIGURECAPTIONS}{FIGURECAPTIONS}}\list
        {Fig. \arabic{enumi}:\hfill}{\settowidth\labelwidth{Fig. 999:}
        \leftmargin\labelwidth
        \advance\leftmargin\labelsep\usecounter{enumi}}}
 \relax
\def\tablecap{\section*{Table Captions\markboth
        {TABLECAPTIONS}{TABLECAPTIONS}}\list
        {Table \arabic{enumi}:\hfill}{\settowidth\labelwidth{Table 999:}
        \leftmargin\labelwidth
        \advance\leftmargin\labelsep\usecounter{enumi}}}
 \relax
 \relax

\catcode`\@=11

\def\marginnote#1{}
\newcount\hour
\newcount\minute
\newtoks\amorpm
\hour=\time\divide\hour by60
\minute=\time{\multiply\hour by60 \global\advance\minute by-
\hour}
\edef\standardtime{{\ifnum\hour<12 \global\amorpm={am}%
    \else\global\amorpm={pm}\advance\hour by-12 \fi
    \ifnum\hour=0 \hour=12 \fi
    \number\hour:\ifnum\minute<100\fi\number\minute\the\amorpm}}
\edef\militarytime{\number\hour:\ifnum\minute<100\fi\number\minute}
\def\draftlabel#1{{\@bsphack\if@filesw {\let\thepage\relax
  \xdef\@gtempa{\write\@auxout{\string
    \newlabel{#1}{{\@currentlabel}{\thepage}}}}}\@gtempa
    \if@nobreak \ifvmode\nobreak\fi\fi\fi\@esphack}
     \gdef\@eqnlabel{#1}}
\def\@eqnlabel{}
\def\@vacuum{}
\def\draftmarginnote#1{\marginpar{\raggedright\scriptsize\tt#1}}
\def\draft{\oddsidemargin -.5truein
        \def\@oddfoot{\sl preliminary draft \hfil
        \rm\thepage\hfil\sl\today\quad\militarytime}
        \let\@evenfoot\@oddfoot \overfullrule 3pt
        \let\label=\draftlabel
        \let\marginnote=\draftmarginnote

\def\@eqnnum{(\theequation)\rlap{\kern\marginparsep\tt\@eqnlabel}%
\global\let\@eqnlabel\@vacuum}  }
\def\preprint{\twocolumn\sloppy\flushbottom\parindent 1em
        \leftmargini 2em\leftmarginv .5em\leftmarginvi .5em
        \oddsidemargin -.5in    \evensidemargin -.5in
        \columnsep 15mm \footheight 0pt
        \textwidth 250mmin      \topmargin  -.4in
        \headheight 12pt \topskip .4in
        \textheight 175mm
        \footskip 0pt

\def\@oddhead{\thepage\hfil\addtocounter{page}{1}\thepage}
        \let\@evenhead\@oddhead \def\@oddfoot{} \def\@evenfoot{}
}
\def\titlepage{\@restonecolfalse\if@twocolumn\@restonecoltrue\onecolumn
     \else \newpage \fi \thispagestyle{empty}\c@page\z@
        \def\thefootnote{\fnsymbol{footnote}} }
\def\endtitlepage{\if@restonecol\twocolumn \else  \fi
        \def\thefootnote{\arabic{footnote}}
        \setcounter{footnote}{0}}  
\catcode`@=12
\relax

\def\ps@headings{\def\@oddfoot{}\def\@evenfoot{}
\def\@oddhead{\hbox{}\hfill
        \makebox[.5\textwidth]{\raggedright\ignorespaces --\thepage{}--
        \hfill }}
\def\@evenhead{\@oddhead}
\def\subsectionmark##1{\markboth{##1}{}}
}

\relax

\def\firstpage#1#2#3#4#5#6{
\begin{document}
\begin{titlepage}
\nopagebreak
\title{\begin{flushright}
         \vspace*{-3cm}
        {\normalsize hep-th/0606051}\\[5mm]
\end{flushright}
\vspace{3cm}
{#3}}
\author{\large #4 \\[0.0cm] #5}
\maketitle
\vskip 2mm
\nopagebreak
\begin{abstract}
{\noindent #6}
\end{abstract}
\vfill
\begin{flushleft}
\rule{16.1cm}{0.2mm}\\[-1mm]
$^\ast$e-mail: kehagias@cern.ch \\[2mm]
June 2006
\end{flushleft}
\thispagestyle{empty}
\end{titlepage}}

\def\simlt{\stackrel{<}{{}_\sim}}
\def\simgt{\stackrel{>}{{}_\sim}}
\newcommand{\dal}{\raisebox{0.085cm}
{\fbox{\rule{0cm}{0.07cm}\,}}}

\newcommand{\be}{\begin{eqnarray}}
\newcommand{\ee}{\end{eqnarray}}
\newcommand{\G}{\Gamma}
\newcommand{\g}{\gamma}
\renewcommand{\d}{\partial}
\newcommand{\bb}{\bar{b}}
\newcommand{\ba}{\bar{a}}
\renewcommand{\a}{\alpha}
\newcommand{\bz}{\bar{z}}
\newcommand{\bt}{\bar{\tau}}
\newcommand{\e}{\epsilon}
\newcommand{\ot}{\otimes}
\newcommand{\p}{\partial}
\newcommand{\btau}{\bar{\tau}}
\renewcommand{\tt}{{\large{\Theta}}}
\newcommand{\bp}{\bar{\partial}}
\newcommand{\cR}{{\cal R}}
\newcommand{\tR}{\tilde{R}}
\newcommand{\tcR}{\tilde{\cal R}}
\newcommand{\hR}{\hat{R}}
\newcommand{\rt}{{{\rho_{\!}}_{}}_{{\tiny  \Theta}}}
\newcommand{\pt}{{{p_{\!}}_{}}_{\Theta}}
\newcommand{\hcR}{\hat{\cal R}}
\newcommand{\oE}{\stackrel{\circ}{E}}
\renewcommand{\p}{\partial}
\renewcommand{\bp}{\bar{\partial}}
\newcommand{\bP}{{\bf P}}
\newcommand{\cn}{{\stackrel{\circ}{\nabla}}}

\newcommand{\gsi}{\,\raisebox{-0.13cm}{$\stackrel{\textstyle
>}{\textstyle\sim}$}\,}
\newcommand{\lsi}{\,\raisebox{-0.13cm}{$\stackrel{\textstyle
<}{\textstyle\sim}$}\,}
\date{}
\firstpage{3118}{IC/95/34}
{\large  {\Large N}EW {\Large T}YPE   {\Large S}CALAR {\Large F}IELDS
{\Large F}OR
{\Large C}OSMIC {\Large A}CCELERATION
\\
\phantom{X}}
{Alex Kehagias$^\ast$}
{\vspace{-.2cm}
\normalsize\sl Department of Physics, National Technical University of Athens,\\
\normalsize\sl GR-15773, Zografou, Athens, Greece } {We present a
model where a non-conventional scalar field may act like dark
energy leading to cosmic acceleration. The latter is driven by an
appropriate  field configuration, which
 result in
an effective  cosmological constant.
The potential role of such a scalar in the cosmological constant problem is also discussed.
}
\newpage

During the last years, evidence of an accelerating expansion of the Universe, first published in~\cite{Perlmutter:1998np},\cite{Riess:1998cb},
 has become one of the most striking results in modern cosmology and
in physics, in general. The most obvious cause of this acceleration is a cosmological constant, although a large number
of alternatives have been proposed~\cite{Copeland:2006wr}. Here we will discuss a rather simple,
although non-conventional,  scalar model, which may also
account for the present Universe acceleration.

The scalar field we will consider is a real one and its dynamics  is
 determined by the action
\be
S=\int d^4x \sqrt{-g}\left(-\frac{1}{2}\d_\mu \Phi(x)\,\d^\mu \Phi(x)
+k\, \d_\mu\Phi(x)\,\d^\mu\Phi(x\!+\!\alpha)\right)\, . \label{action}
\ee
The parameter k is a numerical constant and
the shifts $(x+\a)$ in the argument of the scalar field are generated by a 4-vector $\a^\mu$, which has only a time component,
i.e.,  $\a^\mu=(\a,0,0,0)$.
Hence,  for $x^\mu=(t,x_1,x_2,x_3)$ we have
$(x+\a)=(t+\a,x_1,x_2,x_3)$.
We  assumed for simplicity that there is  no potential for the scalar field.
The unusual feature of the action (\ref{action}) is that, besides the standard kinetic term,
there is also an  additional non-local derivative term~\footnote{This modifies the
 propagator of the scalar field, which for the action (\ref{action}) is given by
\be
\Delta=\Big{\{}p^2\big{[}1-2k \cos(p^\mu\a_\mu)\big{]}\Big{\}}^{-1}\, .
\ee
For $|k|<1/2$ it has a single pole at $p^2=0$.}.
A similar term has been discussed first time to my knowledge,
by Feynman in his thesis~\cite{Feynman:1942us} in a classical mechanical context,
and in connection to the theory of action at a distance. It can be
considered as describing a field interacting with
itself in a distant
mirror by means of advanced and retarded waves. In a modern picture, modelling our universe as a 3-brane,
it could be the result of
a scalar interacting  in a  distant 3-brane through bulk waves. Alternatively, expanding for $t>>a$, we see that
\be
S=\int d^4x \sqrt{-g}\left(-(\frac{1}{2}\!-\!k)\d_\mu \Phi(x)\,\d^\mu \Phi(x)+
k\sum_{n=2}\frac{1}{n!}\d_\mu\Phi(x)a^{\mu_1}...a^{\mu_n}\d^{\mu_1...\mu_n}\Phi(x)\right)\, , \label{action}
\ee
and for $k<1/2$, we have  a higher-derivative theory.
Irrespectively of its origin,
we would like to couple this scalar field to gravity and find
cosmological solutions for the corresponding gravity-scalar system.

The scalar field equation can be obtained from the action (\ref{action}),
after scalar   variations $\delta\Phi(x)$, and  the resulting equation is
\be
\nabla^2\Phi(x)-k \nabla^2\Phi(x\!+\!\a)-k\nabla^2\Phi(x\!-\!\a)=0\, .\label{sca}
\ee
Similarly, the induced energy-momentum tensor is given by
\be
&&T^{(\Phi)}_{\mu\nu}=\,\,\phantom{\frac{1}{2}}\!\!\!\!\!\d_\mu\Phi(x)\,\d_\nu\Phi(x)-
k\,\d_\mu\Phi(x)\, \d_\nu\Phi(x\!+\!\a)-k\,\d_\mu\Phi(x\!+\!\a)\,\d_\nu\Phi(x)
\nonumber \\
&&\hspace{1cm} -\frac{1}{2}g_{\mu\nu}\left[\d_\rho\Phi(x)\, \d^\rho\Phi(x)-2k\,
\d_\rho\Phi(x)\,\d^\rho\Phi(x\!+\!\a)\right]\, , \label{T}
\ee
and, according to the standard prescription, gravity should coupled to it. However, this is not possible in
the present case
and $T^{(\Phi)}_{\mu\nu}$  cannot be the right-hand side
of Einstein equations.
The reason is that, due to the second term
in the action(\ref{action}), general coordinate invariance is lost and consequently, $T^{(\Phi)}_{\mu\nu}$
is not conserved. Indeed,
 it can easily be verified that the  covariant divergence of $T^{(\Phi)}_{\mu\nu}$ does not vanish.
Instead
we find that
\be
\nabla_\mu {{T^{(\Phi)}}^\mu}_\nu=k\nabla^2\Phi(x\!-\!\a)\d_\nu\Phi(x)-k \nabla^2\Phi(x)\d_\nu\Phi(x\!+\!\a)\, .
\ee
 Gravity must couple then to an ``improved"
energy-momentum tensor $\Theta_{\mu\nu}$, which we define as
\be
\Theta_{\mu\nu}=T^{(\Phi)}_{\mu\nu}+t_{\mu\nu}\, ,
\ee
and it is such that $\nabla_\mu\Theta^{\mu\nu}=0$.
Hence, the extra term  $t_{\mu\nu}$ must satisfy
\be
\nabla_\mu {t^\mu}_\nu=k \nabla^2\Phi(x)\d_\nu\Phi(x\!+\!\a)-k\nabla^2\Phi(x\!-\!\a)\d_\nu\Phi(x)\, , \label{t}
\ee
in order $\Theta_{\mu\nu}$ to have vanishing covariant divergence.
Einstein equations are then written  as
\be
R_{\mu\nu}-\frac{1}{2}g_{\mu\nu}R=\kappa^2\Theta_{\mu\nu}\, , \label{ein}
\ee
which is now a consistent system to be solved.

We will look now for cosmological solutions to eqs(\ref{sca},\ref{ein}) of the form
\be
ds^2=-dt^2+a(t)^2\left(dx^2+dy^2+dz^2\right)\, , ~~~~\Phi=\Phi(t)\, .
\ee
Then, we get from eq.(\ref{T}) that the induced energy-momentum tensor is
\be
{{T^{(\Phi)}}^\mu}_\nu={\rm diag}(-\rho_\Phi,p_\Phi,p_\Phi,p_\Phi)
\ee
where
\be
\rho_\Phi=\frac{1}{2}\dot{\Phi}(t)^2- k\dot{\Phi}(t)\dot{\Phi}(t\!+\!\alpha)\, , ~~~~ p_\Phi=\rho_\Phi\, .\label{r}
\ee
Moreover, defining
\be
{t^\mu}_{\nu}={\rm diag}(-\rho,p,p,p)\, ,
\ee
we find that $t_{\mu\nu}$ satisfies eq.(\ref{t}) and $\Theta_{\mu\nu}=T^{\Phi}_{\mu\nu}+t_{\mu\nu}$ is conserved for
\be
\rho&=&k\int_t^{t\!+\!\a}\ddot{\Phi}(\tau\!-\!\a)\dot{\Phi}(\tau)d\tau \, ,\nonumber  \\
p&=&k\dot{\Phi}(t\!+\!\a)\dot{\Phi}(t)-k
\dot{\Phi}(t\!-\!\a)\dot{\Phi}(t)-k\int_t^{t\!+\!\a}\ddot{\Phi}(\tau\!-\!\a)\dot{\Phi}(\tau)d\tau\, . \label{p}
\ee
Indeed, the conservation of $\Theta_{\mu\nu}$ is equivalent to
\be
\dot{\rho}_\Phi+\dot{\rho}+3\frac{\dot{a}}{a}\left(\rho_\Phi+\rho+p_\Phi+p\right)=0\, ,
\ee
which can easily be verified for $\rho_\Phi,p_\Phi,\rho,p$ of eqs.(\ref{r},\ref{p}).
Then, the improved energy-momentum tensor $\Theta_{\mu\nu}$ has the form
\be
\Theta^\mu_\nu={\rm diag}(-\rt,\pt,\pt,\pt)\, ,
\ee
where
\be
\rt\!\!&\!\!=\!\!&\!\!\rho_\Phi+\rho=\frac{1}{2}\dot{\Phi}(t)^2- k\dot{\Phi}(t\!+\!\alpha)\dot{\Phi}(t)
+k\int_t^{t\!+\!\a}\ddot{\Phi}(\tau\!-\!\a)\dot{\Phi}(\tau)d\tau
\,, \nonumber \\
 ~~~\pt\!\!&\!\!=\!\!&\!\!p_\Phi+p=\frac{1}{2}\dot{\Phi}(t)^2-k\dot{\Phi}(t\!-\!\a)\dot{\Phi}(t)
 -k\int_t^{t\!+\!\a}\ddot{\Phi}(\tau\!-\!\a)\dot{\Phi}(\tau)d\tau
\ee
and Einstein  (\ref{ein}) and  scalar equations (\ref{sca}) are explicitly written as
\be
&&\left(\frac{\dot{a}}{a}\right)^2=\frac{\kappa^2}{3}\rt
\, , \label{fre}\\
&&-2\frac{\ddot{a}}{a}-\frac{\dot{a}^2}{a^2}=\kappa^2\pt
\, , \label{fre2}\\
&& \ddot{\Phi}(t)-k\, \ddot{\Phi}(t\!+\!\a)-k\, \ddot{\Phi}(t\!-\!\a)+\nonumber \\
&&+3\frac{\dot{a}}{a}\left(
\dot{\Phi}(t)-k\, \dot{\Phi}(t\!+\!\a)-k\, \dot{\Phi}(t\!-\!\a)\right)=0 \, . \label{sca2}
\ee
A first integral of the scalar equation (\ref{sca2}) is
\be
\dot{\Phi}(t)-k\, \dot{\Phi}(t\!+\!\a)-k\, \dot{\Phi}(t\!-\!\a)=c \, a(t)^{-3} \, , \label{1sca}
\ee
where c is an integration constant.
As usual, only two of the  equations (\ref{fre},\ref{fre2},\ref{sca2}) are independent.
We take Friedman equation (\ref{fre}) and the integrated scalar equation (\ref{1sca}).
With a vanishing $c(=0)$, the solution in the usual case $k=0$ is $\Phi={\rm constant}$. However,
in the present case where  $k\neq 0$, other
solutions exist as well. Indeed, one may easily verify that a solution to eq.(\ref{1sca}) with $c=0$ is
provided by the configuration
\be
\Phi(t)=m\, \Big{(}\cosh\left(\mu \,t\right)+\nu \sinh\left(\mu \,t\right)\Big{)},
 ~~~\mu=\alpha^{-1} {\rm arccosh}(\frac{1}{2k}) \label{sols}
\ee
where $m,\mu$ are mass parameters and $\nu$ is a numerical one.
Then, the right-hand side of Friedman equation (\ref{fre})  turns out to be
\be
\rt =\frac{m^2\sqrt{1-4k^2}}{4 a^2}(1-\nu^2)\, {\rm arccosh}(\frac{1}{2k})^3\, ,
\ee
which is a real constant for $0<k<1/2$. Moreover, we also get that
\be
\pt=-\frac{m^2\sqrt{1-4k^2}}{4 a^2}(1-\nu^2)\, {\rm arccosh}(\frac{1}{2k})^3=-\rt\, ,
\ee
which is the typical equation of state for cosmological constant.
Thus, in this model, the scalar leads to an  effective
 cosmological constant $\Lambda_\Phi$ with value
\be
\Lambda_\Phi=\frac{\kappa^2m^2\sqrt{1-4k^2}}{4 a^2}(1-\nu^2)\,{\rm arccosh}(\frac{1}{2k})^3\, .
\ee
The parameter $\mu$ is of  the order of $\a^{-1}$, and the observed
 cosmic acceleration could be driven by a field like the present one for
 $m^2\approx 10^9 \a^2(1-\nu^2)^{-1}{\rm cm}^{-4}$. Thus, a scalar field with
 dynamics described by the action (\ref{action}) is a
candidate for dark energy. Moreover, it can also be considered, in principle,  as an inflaton field for an early time
 inflation.

We should note here that our scalar field may also be used for the cosmological
constant problem~\cite{Weinberg:1988cp}. In the presence of a  bare
cosmological constant $\Lambda$, Einstein equations (\ref{ein}) are written as
\be
R_{\mu\nu}-\frac{1}{2}g_{\mu\nu}R+\Lambda g_{\mu\nu}=\kappa^2T_{\mu\nu}\, ,\label{einL}
\ee
where $T_{\mu\nu}$ is the matter energy-momentum tensor. With a non-zero $\Lambda$,
flat space-time is not a solution, namely, $g_{\mu\nu}=\eta_{\mu\nu}$ does not satisfy
(\ref{einL}). However, for $T_{\mu\nu}=\Theta_{\mu\nu}$ the bare cosmological
constant $\Lambda$  is shifted to $\Lambda+\Lambda_\Phi$   by the scalar field (\ref{sols}) and it can take any
value  by appropriate choice
of the numerical parameter $\nu$. Then, by adjusting $\nu$ so that $\Lambda_\Phi=-\Lambda$, flat space-time is now
a solution to eq.(\ref{einL}). This is always possible for a negative bare cosmological constant,
whereas, for a positive one, the parameter $\nu$ should be such that $\nu>1$. In the latter case,
although $\rt$ is negative violating seemingly weak energy conditions, the energy density $\rho_\Phi$
of the scalar field associated with  the canonical energy-momentum tensor is positive for $t>0$.
In other words,
for any value of
$\Lambda$, there always exists a scalar described by
the action (\ref{action}) and profile given by (\ref{sols}),
 which may nullify
$\Lambda$ or drive it to its present  value. The question why this value is so tiny remains open.

\vskip .2in

\noindent {\bf Acknowledgement.} This work is supported by the EPEAEK programme ``Pythagoras" and co-funded by
the European Union (75\%) and the Hellenic State (25\%).

\end{document}